\documentclass[runningheads]{llncs}

\usepackage{cite}
\usepackage{amsmath,amssymb,amsfonts}
\usepackage{tabularx}
\usepackage{algorithmic}
\usepackage{graphicx}
\usepackage{textcomp}
\usepackage{supertabular}
\usepackage{longtable}
\usepackage{geometry}
\usepackage{caption}
\usepackage{enumitem}
\usepackage{array}
\usepackage{ragged2e}

\usepackage{xcolor}
\usepackage{rotating}

\usepackage{geometry}
\usepackage{enumitem}
\twocolumn

\def\BibTeX{{\rm B\kern-.05em{\sc i\kern-.025em b}\kern-.08em
    T\kern-.1667em\lower.7ex\hbox{E}\kern-.125emX}}

\begin{document}
\title{Machine Unlearning for Recommendation Systems: An Insight\\
{\footnotesize}}

\titlerunning{Machine Unlearning for Recommendation Systems}

\author{Bhavika Sachdeva\thanks{These authors should be considered co-first authors.}\inst{1} \and
Harshita Rathee\textsuperscript{*}\inst{1} \and Sristi\textsuperscript{*}\inst{1} \and  Arun Sharma\inst{1}
\and
Witold Wydmański\inst{2}}

%
\authorrunning{B. Sachdeva et al.}
%
\institute{Department of Computer Science and Engineering,
Indira Gandhi Delhi Technical University for Women, New Delhi, India\\ \email{bhavika2210@gmail.com}, \email{harshita10.17@gmail.com}, \email{sristi0108@gmail.com} and \email{arunsharma@igdtuw.ac.in} \and Faculty of Mathematics and Computer Science, Jagiellonian University, Kraków, Poland
\email{witold.wydmanski@uj.edu.pl}}

\maketitle

\begin{abstract}

This review explores machine unlearning (MUL) in recommendation systems, addressing adaptability, personalization, privacy, and bias challenges. Unlike traditional models, MUL dynamically adjusts system knowledge based on shifts in user preferences and ethical considerations. The paper critically examines MUL's basics, real-world applications, and challenges like algorithmic transparency. It sifts through literature, offering insights into how MUL could transform recommendations, discussing user trust, and suggesting paths for future research in responsible and user-focused artificial intelligence (AI). The document guides researchers through challenges involving the trade-off between personalization and privacy, encouraging contributions to meet practical demands for targeted data removal. Emphasizing MUL's role in secure and adaptive machine learning, the paper proposes ways to push its boundaries. 
The novelty of this paper lies in its exploration of the limitations of the methods, which highlights exciting prospects for advancing the field. 
\end{abstract}

\section{Introduction}

Machine learning (ML) stands as a transformative force, revolutionizing data processing and analysis across diverse domains. Its pervasive impact on our approach to data is particularly evident in recommendation systems, a critical subset of ML that significantly influences our digital experiences.

Within the intricate realm of recommendation systems, these algorithms are architects of personalized user interactions. Operating on sophisticated algorithms and data-driven insights, these systems are crucial in delivering tailored content and product suggestions. Through a nuanced understanding of user behaviors and preferences, recommendation systems contribute to an enriched online experience by seamlessly integrating relevant and engaging recommendations.

At the core of this intricate interplay lies the pivotal role of ML, serving as the backbone that propels the nuanced mechanics of recommendation systems. These systems, meticulously designed to decipher and respond to user preferences and behaviors, exemplify the symbiotic relationship between human interaction and algorithmic intelligence. The synergy between ML and recommendation systems highlights the depth of their impact on how we consume and engage with content in the digital realm.

As we delve further into the intricacies of recommendation systems, it becomes apparent that machine learning's role is not ancillary but foundational. Its integration empowers these systems to continually evolve and adapt, fine-tuning their recommendations based on the dynamic landscape of user preferences. This symbiotic relationship underscores the dynamic nature of the intersection between machine learning and recommendation systems, marking it as a compelling frontier in the evolving landscape of digital technology.
These applications vary from news portals, social media\cite{socialmedia1}, \cite{socialmedia2}, e-commerce platforms to OTT platforms such as Netflix\cite{netflix}. Random walk-based models, such as Pixie \cite{Pixie}, have been successfully employed in large-scale industrial contexts to deliver highly effective personalized recommendations. There are several types of filtering algorithms used like Collaborative filtering, Content based filtering, and matrix factorization. Many of the recommendation systems use hybrid algorithms which combine several recommendation techniques into one. Once created, these recommendation systems have the capacity to retain knowledge from their training data.

However, with the ever-growing importance and prevalence of recommendation systems, a challenging concept looms in the background – the need for machine unlearning. As these systems continuously evolve and adapt, there arises a crucial question: how do we effectively mitigate or alter the impacts of prior learning, both within the data repositories and the core ML models that power recommendation systems? 

MUL is an emerging science that tries to address the complexities associated with undoing or revising the consequences of prior learning, both within the data repositories and the underlying machine learning models. It arises from the recognition that, in a constantly evolving and privacy-conscious digital landscape, users may require their data to be selectively erased or modified, not only to enhance privacy and data security but also to maintain the accuracy and relevance of recommendations.

However, an essential consideration arises in this context \cite{MUL}: throughout the life cycle of recommendation algorithms under examination, instances arise where it becomes imperative for the system to selectively discard specific information and its complete historical context.

Given that all these algorithms rely on the premise of retaining data, a significant and pressing concern associated with these systems is the potential compromise of user privacy. Recent studies have brought to light that trained models, including recommender systems, large pre-trained models, and fine-tuned natural language models, can inadvertently expose sensitive user information. In such cases, users are actively looking for methods to remove the impact of their sensitive data from these models.  

Another important rationale for the need for recommendation unlearning pertains to the system's utility. Over time, continuous data collection is essential for enhancing existing models. However, some of this newly acquired data can be harmful, such as tainted data resulting from poisoning attacks or data that diverges from the model's typical distribution (out-of-distribution data)\cite{socialmedia2}. This subpar data can have a notable detrimental impact on the effectiveness of recommendations. Once identified, it becomes imperative for the system to expunge such data to restore its usefulness.  

Furthermore, in our ever-evolving world, users increasingly seek recommendations that adapt to their changing needs. Picture a scenario involving political posts—a user was once actively engaged in such content but has since lost interest. Despite this shift, the user continues to receive recommendations related to political discourse. This example underscores the need for users to have control over the model's historical data, particularly in the realm of political content, to ensure that recommendations remain relevant and responsive to their evolving preferences and interests. 

The most basic approach to unlearning entails retraining the model using its original data, but this method presents a substantial challenge due to the significant computational demands, especially for large-scale datasets. For recommendation unlearning to be successful, it must not only involve purging collected user data but also erasing the influence of that data on the model. This process should adhere to three fundamental principles: completeness, utility, and efficiency\cite{matrix}. One direct method to satisfy both completeness and utility is to retrain the model entirely from the ground up. However, this approach is often impractical due to the high costs associated with retraining, rendering it inefficient for real-world applications. To ensure the efficiency of the unlearning process, existing methods for recommendation unlearning have had to strike a balance between the three fundamental principles, considering the trade-offs between these factors and efficiency.

Concerning recommendation systems, the existing MUL methods face challenges due to the complex geometric relationships and structures inherent in the data they handle. In addition to exploring the traditional recommendation unlearning algorithm, this paper places emphasis on graph-based unlearning algorithms.\cite{gnndelete}. In the quest for refining recommendation systems, a significant aspect that has garnered attention is the challenge posed by evolving user preferences. Therefore, the adaptability of recommendation systems to capture and respond to these nuances becomes paramount. Addressing this challenge involves exploring novel approaches to real-time learning and dynamic model adjustments. 

The study's main contributions include thoroughly examining the concept of machine unlearning (MUL) within recommendation systems. It offers a deep dive into the adaptable nature of MUL, focusing on its implications for personalization, privacy, and bias – all critical aspects of today's AI-driven systems. The research provides a nuanced perspective on how MUL could potentially revolutionize recommendation systems by critically evaluating its fundamentals, its practical applications, and the complex issue of algorithmic transparency.

Additionally, the paper highlights the delicate balance between tailoring recommendations to individual preferences and respecting user privacy. It emphasizes the importance of building user trust in these systems and suggests avenues for future research that prioritize responsible and user-centric artificial intelligence.

Through its proposals to expand secure and adaptive machine learning using MUL, this study significantly contributes to ongoing discussions about ethical practices in AI. Its unique exploration of MUL's transformative capabilities within recommendation systems offers valuable insights into ethical considerations and advances the conversation on responsible AI development.

The structure of this paper is as follows: Section 2 discusses the review methodology used for the paper. Section 3 provides a thorough literature review of existing state-of-the-art (SOTA) algorithms on machine unlearning for recommendation systems. Section 4 summarizes the discussions and future directions of the reviewed methodologies in the relevant field, and finally, Section 5 details the conclusion.

\section{Review Methodology}
In conducting this literature review, a systematic approach was adopted to identify relevant sources and contributions. Thorough searches were conducted across esteemed academic databases, including IEEE journals, Elsevier publications, and Springer databases. The selected keywords employed in the search process encompassed terms such as "machine unlearning," "recommendation systems," "graph networks," "recommendation unlearning methods," and "privacy in machine learning." A representative selection of the most pertinent and recent research in machine unlearning within recommendation systems was compiled by employing these comprehensive search strategies and leveraging authoritative databases.

\section{Literature Review}
This section takes a deep dive into the ever-evolving realm of machine unlearning in recommendation systems, specifically focusing on the utilization of graph networks. It sheds light on several seminal contributions in this nascent field, each presenting a distinct approach to the intricate task of unlearning within recommendation systems. These contributions offer insights into various techniques and strategies, encompassing advanced data partitioning and adaptive aggregation methods, novel recommendation unlearning methodologies, and model-agnostic layer-wise operators.  There exists scope for extensive work in the field of machine unlearning, particularly in the domain of graph networks primarily used by recommendation systems. Some significant contributions include:

\subsection{RecEraser}

The authors introduced RecEraser\cite{receraser}, an efficient machine unlearning framework that can be utilized for recommendation systems.The core concept behind RecEraser involved segmenting the training data into multiple shards and subsequently training a separate model for each shard. In order to maintain collaborative information within the data, 3 innovative data partition algorithms were initially developed to create balanced groups based on their similarity. Additionally, they introduced an adaptive aggregation method to enhance the effectiveness of the global model while preserving collaborative insights. RecEraser aims to protect and harness collaborative information by carefully partitioning the dataset while making shards. It adaptively assigned weights to these shards at the time of aggregation.\cite{survey}

However, RecEraser does not consider the transport weights between sub-models, which may lead to sub-optimal clustering and imbalance of group data. \cite{ultrare} It does not guarantee the convergence of sub-models on their own group data, which may affect the quality of recommendations. RecEraser uses a complex combination estimator that requires additional training and tuning, which may increase the computational cost and risk of overfitting. \cite{matrix}

\subsection{Recommendation Unlearning via Matrix Correction}
MCRU\cite{matrix} discussed a novel approach to recommendation unlearning, which is the process of deleting specific data and models from a trained recommender system. The authors introduced Interaction and Mapping Matrices Correction (IMCorrect), a technique for recommendation unlearning. IMCorrect was able to attain recommendation unlearning through updating the interaction matrix and improving the completeness as well as utility. This was done by updating the mapping matrix without expensive compute process of retraining the model. IMCorrect is a whitebox model that delivered more significant flexibility in managing diverse recommendation unlearning plans. It has the distinctive ability of learning incrementally from unknown data, which further improved its usefulness.\cite{matrix2}

\subsection{AltEraser}
AltEraser\cite{alteraser} is an important machine unlearning work for real-world applications where users desired that a portion of their data be removed, not exclusively from the data storage but correspondingly from the ML models being used in the process, both for privacy or utility grounds. The authors explored swift ML techniques for recommendation engines that could dismiss the impact of a short portion of the training set from the recommendation system without the full cost that would be incurred upon retraining. A realistic method to accelerate this was to fine-tune the existing recommendation system on the left training set rather than initiating from a point of random initialization. The authors presented a new recommendation unlearning strategy, AltEraser\cite{alte2}, which separates the optimization issue of unlearning and divides it additionally into tractable sub-problems.

\subsection{GNNDelete}

GNNDelete framework\cite{gnndelete} introduced an innovative strategy for graph unlearning within Graph Neural Networks (GNNs). It involved the removal of elements such as nodes, their labels, and relationships from a pre-trained GNN model, a critical process for practical applications where data privacy, accuracy, and relevance are paramount concerns. This model-agnostic approach prioritized two key attributes for effective graph unlearning: Neighbourhood Influence and Deleted Edge Consistency. The layer-wise operator employed in GNNDelete optimized these qualities, ensuring that model weights and neighboring representations remain uninfluenced by deleted elements. The Deleted Edge Consistency guaranteed the removal of any residual impact, while Neighbourhood Influence preserved the integrity of the remaining model knowledge post-deletion. GNNDelete achieved the removal of nodes and edges from the model while retaining crucial learned information by modifying representations. Notably, this approach addressed the limitations of current partitioning and aggregation-based techniques, particularly in handling local graph relationships.\cite{distil}

\subsection{Caboose Forget Me Now: Fast and Exact Unlearning in Neighborhood-based Recommendation}
The authors\cite{forget} proposed an algorithm that leveraged the data's sparsity to work only using top-k entries that are presently impacted by the unlearning process. This meant that it could be computed using the norm of vectors and dot products. It also concentrated on zero similarity functions when the dot product was zero between two rows. This helped to avoid unnecessary computations for pairs of users that hadn't engaged with any shared items. The algorithm's individual stages were parallelized \cite{forget2} to use multiple processing units for faster computation. Appropriate data structures, such as compressed representations for binary heaps as well as sparse matrices for managing top-k lists, were chosen to optimize low-level operations in the algorithm.
 
\subsection{Graph Scattering Transform (GST) Unlearning Graph Classifiers with Limited Data Resources}

The method's \cite {gst} computational complexity for node removal was considered order-optimal with respect to unlearning complexity for graph classifiers. This efficiency was a significant advantage, especially compared to methods that frequently required full retraining. The approach's use of GSTs for different training graphs resembled sharding with small components. Notably, the sizes of these shards did not impact the performance of the final model, providing a robust and flexible system for graph classification tasks. The authors suggested exploring other loss functions or methods in specific applications was one potential future direction. This indicated the possibility of extending the framework's capabilities even further. The first nonlinear graph learning framework based on GSTs with an approximate unlearning mechanism offered versatility, efficiency, and certified removal with lower privacy costs than alternatives. Especially for limited data resources, these methods have been proven effective.\cite{limited}

\subsection{Selective and Collaborative Influence Function (SCIF) for Efficient Recommendation Unlearning}

SCIF\cite{selective} was designed to enhance unlearning performance in recommendation tasks by incorporating two critical components: user selection and preservation of collaboration. Each data point was treated independently in the the traditional IF-based unlearning method. Thus, ignoring the collaborative nature of recommendation data, SCIF introduced a collaborative component to the influence function. When unlearning a data point, this component restored collaboration among users and items. Calculating the influence on the target user embedding can still be computationally demanding, even with the reduction in parameters considered. To address this, authors employed techniques like Hessian-Vector Product (HVP) and conjugate gradients, which made the computation more efficient. Neural Matrix Factorization (NMF) and LightGCN were the recommender models used.

\subsection{Attribute-wise Unlearning in Recommender Systems}
Attribute-based unlearning methods for recommendation systems were also developed to protect the sensitive attributes of users. This was done by using Post-Training Attribute Unlearning.\cite{selective2}. The collaborative filtering technique that is most commonly used for recommendations was selected in this study. It is based on matrix factorization and was used to split the user-item interaction matrix into 2 low-rank embedding matrices. These are item and user embedding. The attribute with two or binary labels was the primary focus of the work.

\subsection{Recommendation Unlearning via Influence Function (IFRU)}

IFRU\cite{ifru} leveraged influence functions, a concept from differential privacy and sensitivity analysis. Influence functions(IF) provided a way to gauge the impact of individual data points on the output provided by the model. Influence functions helped identify and adjust the impact of specific data points, enabling the model to unlearn undesirable or sensitive information without needing full retraining. The key idea behind IFRU was to compute the unlearning of influential data points followed by modifications to the model's parameters. As a result, the impact of those data points on the predictions made by the model was effectively diminished. This allowed the model to "forget" the unwanted information and adapt to the updated training data, ultimately improving the model's safety and privacy. Matrix Factorization(MF) and LighGCN were the recommender models used.\cite{ifru2}

\subsection{Federated Unlearning for On-Device Recommendation}

Federated Recommendation Unlearning (FRU)\cite{fed} efficiently unlearned target users in federated recommendation systems. It calibrated and combined all users' previous model updates to recreate the Federated Recommendation (FedRec)\cite{fedrec} models. Neural collaborative filtering and LightGCN are the FedRec algorithms used. FRU sped up the reconstruction process compared to retraining from scratch as it reliably rolled back FedRec and calibrated the stored model changes. FRU was model-agnostic and was likely utilized in several FedRecs. The findings revealed that FRU could remove specific users' impact and effectively retrieve FedRec's seven times faster. FRU performed unlearning when many users rendered the FedRec service and requested their information be forgotten at a specific time. This involved restoring the FedRec model to an earlier state and calibrating the historical updates of the model on the existing clients. As the direction drove the model fitting, the calibration focused on the direction of updates while maintaining their length unchanged.\cite{fed1}

\subsection{Heterogeneous Federated Knowledge Graph Embedding Learning and Unlearning}

FedLU\cite{hetfed} introduced embedding learning and frameworks based on federated Knowledge Graphs, which showcased innovation, addressing knowledge sharing and forgetting in a federated environment. The model effectively balanced global convergence and local optimization through mutual knowledge distillation, demonstrating a thoughtful approach to federated learning. Retroactive interference and passive decay in federated unlearning allowed the model to forget specific knowledge while maintaining overall performance, showcasing a robust unlearning mechanism. FedLU \cite{hetfed2} has the potential to make a significant impact by addressing challenges in federated KG embedding, including knowledge exchange and unlearning, contributing to advancements in federated learning research.
\subsection{Adv-MultVAE Model: Unlearning Protected User Attributes in Recommendations with Adversarial Training}

The authors acknowledged the limitations of their approach and suggested potential areas for improvement in the future. These included exploring generalization aspects, understanding user perceptions of bias in recommendations, and identifying which user groups are most affected by the approach\cite{AdvMUl}. These suggestions indicated a commitment to ongoing research and improvement.
There was a marginal decrease in performance, mainly attributed to the challenge in model selection. \cite{AdvMUl2}The authors recognized that finding a balance between accuracy (BAcc) and recommendation quality (NDCG) helped mitigate this performance loss. It was essential to consider this trade-off when implementing the approach.

\subsection{MUter: Machine Unlearning on Adversarially Trained Models}

MUter\cite{muter} presented a pioneering approach to the joint challenge of privacy and robustness in unlearning from adversarial training models. It used a Hessian-based measure and efficiency enhancements that marked a significant advancement. The method proved to be versatile and effective across linear and neural network models and opened avenues for future exploration. Overall, MUter stood out as a favorable solution with a dual focus on privacy and adversarial robustness.

\subsection{Laser: Making Recommender Systems Forget: Learning and Unlearning for Erasable Recommendation}

LASER\cite{laser} presented a promising framework for addressing the challenge of unlearning in recommendation systems. The combination of the Group module, focusing on balanced user grouping through collaborative embedding, and the SeqTrain module, implementing a sequential training approach with collaborative cohesion as a difficulty measure, demonstrated an innovative strategy for achieving efficient unlearning and improved model utility. The theoretical support for the SeqTrain module added credibility to the proposed approach, and empirical validation on real-world datasets strengthening its practical relevance. However, potential complexities in implementation, dependency on the quality of collaborative embedding, and sensitivity to hyperparameters should be carefully considered. Overall, LASER provided a valuable contribution to the field, and its real-world implications and scalability in different recommendation system scenarios warrant further exploration and research.

The most recent works on MUL significantly improve existing techniques, such as UltraRE \cite{ultrare}, an ensemble-based framework for recommendation unlearning that addresses redundancy, relevance, and combination losses to enhance model utility. Another notable contribution introduced by the certified MUL algorithm for minimax models \cite{delete}, employs a total-Hessian-based complete Newton update and the Gaussian mechanism from differential privacy. Furthermore, these works, including exploration-focused approaches \cite{explore}, evaluate current unlearning methods and introduce techniques like improved adversarial attacks, with a particular emphasis on recommendation systems.

\subsection{Machine Unlearning from Bi-linear Recommendations}
The work\cite{netflix} proposed a fast heuristic-based MUL technique for recommendation systems, Untrain-ALS, that unlearns a bi-linear model without compromising recommendation accuracy. The paper discussed the privacy risks that bi-linear recommendation systems hold on memorizing data. They effectively presented an empirical test using de-noised membership inference that proved sensitive to bi-linear recommendation engines' memorization.

\onecolumn 

\begin{longtable}{|p{1.75cm}|c|p{2.5cm}|p{6.5cm}|p{5.5cm}|}
\caption{Analysis of SOTA MUL Algorithms on Recommendation Systems}
\\
\hline
\textbf{Method} & \textbf{Year} & \textbf{Datasets} & \textbf{Pros} & \textbf{Limitations} \\
    \hline
    \endfirsthead
    \multicolumn{5}{c}{{\tablename\ \thetable{}: Continued from previous page}} \\
    \hline
    \textbf{Method} & \textbf{Year} & \textbf{Datasets} & \textbf{Pros} & \textbf{Limitations} \\
    \hline
    \endhead
    \hline
    \multicolumn{5}{r}{{Continued on the next page}} \\
    \endfoot
    \hline
    \endlastfoot
    RecEraser\cite{receraser}& 2022& Yelp2018, Movielens-1m, Movielens-10m& The proposed model employed balanced data partition for collaborative information preservation, attention-based adaptive aggregation with self-adaptive learning rates, and computationally efficient unlearning by retraining only relevant submodels and the aggregation part, demonstrating superior performance compared to SOTA algorithms.& Efficiently unlearning employing the Sharded, Isolated, Sliced, and Aggregated (SISA) framework in batch setting remains an issue. For some scenarios, the brute force method of retraining performed better than RecEraser.
    RecEraser does not consider the transport weights between sub-models, which may lead to sub-optimal clustering and imbalance of group data. It does not guarantee sub-models' convergence on their own group data, which may affect the quality of recommendations. It only handles individual forgetting requests, and its efficiency may decrease when dealing with batch requests. \cite{ultrare}
    \\
    \hline
    AltEraser\cite{alteraser}& 2022& MovieLens-1m, Amazon-14core, KuaiRec-binary& AltEraser demonstrated promising results in consistency, accuracy, and efficiency as the foremost try at rapid approximate MUL for neural recommendation models that are SOTA, based on extensive experiments on three real-world datasets& AltEraser, a fine-tuning method proposed in the paper, may be sensitive to hyperparameters like learning rate, and its performance on recommender systems with non-linear models or intricate user-item interaction patterns remains uncertain. \\
    \hline
    MCRU\cite{matrix}& 2023& MovieLens-1m, Gowalla, Yelp&IMCorrect demonstrated superiority in forgetting out-of-distribution, out-of-date, and attack data, exhibiting effectiveness in completeness, utility, and efficiency, and proved to be a versatile tool applicable in various recommendation unlearning scenarios.& IMCorrect assumes correctability of the interaction matrix, which may not be viable when unlearned data significantly impacts the model, and the paper lacks discussion on the method's performance in large-scale recommender systems with millions of users and items. \\
    \hline
    GNNDelete\cite{gnndelete} & 2023 & Cora, PubMed, DBLP, CS, OGB-Collab, OGB-BioKG, WordNet18RR& Regarding edge, node, and nodal feature deletion tasks, GNNDelete performed better than current methods by as much as 38.8\% (AUC) and 32.2\% when differentiating removed edges from non-deleted ones. Compared to retraining the GNN system from WordNet18, it required 9.3x less space and less time than 12.3x, indicating its efficiency& Further exploration is needed to assess GNNDelete's performance on larger datasets and diverse graph types, as well as to investigate its robustness to noise and adaptability to dynamic graphs with constant node and edge changes.\\
    \hline
    Caboose\cite{forget} & 2023 & Movielens-10m, Lastfm, Synthetic Interactions, Yahoo, Spotify& Caboose facilitated quick index building, enabling sub-second unlearning for large datasets and seamless integration with next-basket recommendation models, offering transparent and cost-effective alternatives to neural approaches with significantly reduced training time.& The removals' impact depended on the data's model details and co-occurrence structure.\\
      \hline
   SCIF \cite{selective} & 2023 & Movielens-1m, Amazon Digital Music (ADM)&The method eliminated the need for retraining, making it ideal for large-scale systems, enhancing efficiency through selective user embedding updates, preserving collaboration, and employing a Membership Inference Oracle (MIO) that verified comprehensive unlearning, ultimately demonstrating significant improvements in efficiency while maintaining completeness.& Collaborative filtering encountered challenges with large datasets and required further development for adaptation to complex models by incorporating additional sources of user-item interaction data.\\
      \hline
    GST\cite{gst} & 2022 &  MNIST, CIFAR10, PROTEINS, IMDB, COLLABS & The graph classifier offered mechanism, adaptable to various loss functions, with an approximate removal guarantee within privacy constraints, surpassing DP-GNNs in privacy cost for similar unlearning outcomes for efficient unlearning with nonlinearities in graph embedding and reduced training data. & Comparison with DP-GNNs revealed their focus on node-level privacy in classification tasks, suggesting their suitability for node classification rather than graph classification.\\
      \hline
   IFRU\cite{ifru} & 2023 & Amazon Electronics, BookCrossing& IFRU aims for comprehensive unlearning by removing the influence of unusable data without modifying model architectures, ensuring thoroughness without side effects on recommendation quality. It efficiently addressed unlearning in recommendation systems, especially in scenarios requiring updates based on user interactions or varying sensitivity levels.& IFRU's effectiveness was influenced by recommendation model complexity, potentially facing challenges with extremely complex models. It may also have limitations in highly sparse datasets, particularly when interactions between users and items are limited, making accurate influence function calculation and effective recommendations challenging..\\
    \hline
 FRU\cite{fed} & 2022 & MovieLens-100k, Steam-200k& FRU allowed users to request data erasure for privacy compliance, enabling specific data removal without full model retraining, utilizing efficient storage methods to reduce historical data storage, and achieving quick recalibration for up to 7 times better system responsiveness.& The potential drawbacks involved the complexity of managing updates on a distributed system and the need for additional computational resources for update revision and reconstruction.\\
    \hline
Adv-MultVAE\cite{AdvMUl} & 2022 & MovieLens-1M, LFM-2b-DemoBias & Adv-MultVAE integrated an adversarial component to mitigate societal biases in recommendation systems, effectively reducing biases in latent information about user-protected attributes, promoting fairness and privacy, as validated through empirical evidence measuring both the reduction in encoded protected information and recommendation accuracy.& The adversarial approach in bias reduction lacked effective generalization to new datasets and changing user behavior, and it may not fully address user perception of bias in recommended results, highlighting potential limitations in its application.\\
    \hline
    Laser\cite{laser} & 2022 & MovieLens-1M, Amazon Digital Music  & LASER divided users into balanced groups using collaborative embedding and enhanced retraining efficiency, potentially lowering computational expenses. The SeqTrain module adopted a sequential training method with collaborative cohesion as a measure of difficulty, offering a systematic approach that enhanced model utility. & LASER's effectiveness depends on collaborative embedding quality and sensitivity to hyperparameter choices, impacting its performance.\\
    \hline
   FedLU\cite{hetfed}& 2023 & FB15k-237 C3,C5,C10& FedLU employed mutual knowledge distillation for global convergence and enhanced local optimization on dynamic, heterogeneous data among clients. It integrated retroactive interference and passive decay for effective triplet unlearning without significant performance degradation.& FedLU introduced complexity with mechanisms like mutual knowledge distillation, retroactive interference, and passive decay, demonstrating effectively on specific clustering-based datasets, though performance could have varied on datasets with different characteristics.\\
    \hline
   MUter\cite{muter}& 2023 & MNIST-B, Covtype, Lacuna-10, CIFAR-10& MUter addressed the challenge of unlearning from adversarial training models, ensuring privacy compliance and robustness, utilizing a Hessian-based measure with computational techniques for efficiency gains, demonstrating high effectiveness in maintaining model accuracy and adversarial robustness across linear and neural network models.& While efficient, MUter incurred a slightly longer computational time due to the additional computation of the total Hessian.\\
    \hline

\end{longtable}

\twocolumn 

\subsection{Closed-form Machine Unlearning for Matrix Factorization (CMUMF)}
A closed-form MUL technique is suggested by the authors\cite{matrixF}. As the closed-form unlearning update, the authors recorded the implied reliance between the rows and columns, leading to a complete Newton step based on Hessian. The paper validated the efficacy and utility of CMUMF using 5 real-world datasets from 3 distinguishable domains of application, including artificial data sets with three different sizes.

\section{Challenges and Future Directions}

Machine unlearning is a fairly contemporary and emerging area of research, focusing on the ability to remove or update specific data from machine learning models. As researchers delve into this emerging domain, it becomes evident that exploring unlearning techniques has been somewhat limited, particularly in the context of diverse data structures like multimodal data. Future investigations must extend the scope of unlearning to intricate formats such as text, audio, and multimedia, presenting novel challenges related to temporal sequences, spatial linkages, and hierarchical structures.

While existing research has made strides in understanding unlearning within convex models like logistic regression, there remains a notable gap in addressing the complexities introduced by non-convex neural networks, including Convolutional Neural Networks and Recurrent Neural Networks commonly employed in deep learning applications. Efficient algorithms tailored for non-convex optimization problems in the context of unlearning strategies are still in the early stages of development.

Moreover, the current focus on instance-level data removal in unlearning strategies has limitations, offering users only a constrained degree of flexibility in data management and model updates. Future research should explore the development of quantitative metrics to foster a more versatile and user-friendly approach. These metrics could provide a nuanced evaluation of unlearning efficacy by measuring the extent of influence retention for retained data and the corresponding reduction in influence for removed data. This avenue of exploration promises to enhance the comprehensiveness and adaptability of machine unlearning methodologies, paving the way for future more effective and user-centric approaches.

Future scope for this can be extended to social media recommendation systems where the reliance on user search history is pretty high. Once any information is fed into them, the graphically dense relation that information establishes with the existing information makes the problem more challenging and relevant to the actual problems. Solving these problems is important to ensure future development in this domain.

\section{Conclusion}
In conclusion, this literature review highlighted the notable contributions in this domain, underscoring the adoption of graph networks to address the multifaceted challenges of machine unlearning. Significant advancements and innovative strategies were observed; however, it is imperative to recognize that this domain remained in its premature stages. The niche area of machine unlearning within recommendation systems presented ample opportunities for future exploration and development. Researchers and practitioners were able to harness the potential of machine unlearning to enhance user privacy, maintain recommendation system accuracy, and adapt to ever-changing user preferences. As recommendation systems continued to shape our digital experiences, the quest for efficient, accurate, and privacy-aware MUL methods ensured these systems' ongoing success and safety.


\bibliography{Machine_Unlearning_for_Recommendation_Systems__An_Insight}
\bibliographystyle{IEEEtran}

\end{document}